# Review of the Low-Frequency 1/$f$ Noise in Graphene Devices

## Implications of 1/f Noise for Practical Applications of Graphene

# Alexander A. Balandin

Nano-Device Laboratory, Department of Electrical Engineering and Materials Science and Engineering Program, Bourns College of Engineering, University of California – Riverside, Riverside, California 92521 USA

### Abstract

**Low-frequency noise with a spectral density that depends inversely on frequency ($f$) has been observed in a wide variety of systems including current fluctuations in resistors, intensity fluctuations in music and signals in human cognition. In electronics, the phenomenon, which is known as 1/$f$ noise, flicker noise or excess noise, hampers the operation of numerous devices and circuits, and can be a significant impediment to development of practical applications from new materials. Graphene offers unique opportunities for studying 1/$f$ noise because of its 2D structure and carrier concentration tuneable over a wide range. The creation of practical graphene-based devices will also depend on our ability to understand and control the low-frequency 1/$f$ noise in this material system. Here, I review the characteristic features of 1/$f$ noise in graphene and few-layer graphene, and examine the implications of such noise for the development of graphene-based electronics including high-frequency devices and sensors.**

Low-frequency noise with the spectral density $S(f) \sim 1/f^{\gamma}$ (where $f$ is the frequency and $\gamma \approx 1$) was discovered in vacuum tubes [1] and later observed in a diverse array of systems [2-5]. In electronics, this type of noise, which is commonly referred to as $1/f$ noise, flicker or excess noise, is usually found at $f < 100$ kHz. The corner frequency $f_c$, where the $1/f$ noise level is equal to that of thermal or shot noise, ranges from a few Hz to tens of kHz and is often used as a figure of merit for the $1/f$ noise amplitude. The importance of $1/f$ noise in electronics has motivated numerous studies of its physical mechanisms and the development of a variety of methods for its reduction [6]. However, despite almost a century of research, $1/f$ noise remains a controversial phenomenon and numerous debates continue about its origin and mechanisms.

The general name for this intrinsic noise type does not imply the existence of a common physical mechanism giving rise to all its manifestations [7]. It is now accepted that different fluctuation processes can be responsible for the $1/f$ noise in different materials and devices. For this reason, practical applications of a new material system usually require a thorough investigation of the specific features of the low-frequency noise in the material and the development of methods for their reduction. For example, the introduction of GaN/AlGaN wide-band gap semiconductors into communication technologies relied on reducing the level of $1/f$ noise by about five orders of magnitude, which was achieved through several years of research and development [6, 8].

Fluctuations in the electrical current, $I \propto qN\mu$, can be written as $\delta I \propto q(\delta N)\mu + qN(\delta\mu)$, where $q$ is the charge of an electron, $N$ is the number of charge carriers and $\mu$ is the mobility. Correspondingly, one can distinguish the *mobility fluctuation* and *carrier number fluctuation* mechanism of $1/f$ noise [7]. Box I provides a summary of the intrinsic noise types and theory basics. It is generally accepted that in conventional semiconductor devices such as Si complementary metal-oxide-semiconductor (CMOS) field-effect transistors (FETs), $1/f$ noise is described well by the McWhorter model, which uses the carrier-number fluctuation approach (see Eq. (B1)). In metals, on the other hand, $1/f$ noise is usually attributed to mobility fluctuations. The mobility fluctuations can arise from fluctuations in the scattering cross-section of scattering centers (Eq. (B2)). There are materials and devices where contributions from both mechanisms are comparable or cross-correlated. The location of the noise sources − *surface* vs. *volume* of the electrical conductor − has also been a subject of considerable debates [7, 9-12].

Graphene is a *unique* material system in the 1/*f* noise context owing to its two-dimensional (2D) nature, unusual linear energy dispersion for electrons and holes, zero energy band gap, specific scattering mechanisms, and metallic type conductance. From one side, it is an ultimate surface where conduction electrons are exposed to the traps, e.g. charged impurities in a substrate or environment, which can result in strong carrier number fluctuations. From the other side, graphene can be considered a zero-band-gap metal, where mobility fluctuations owing to the charged scattering centers in the substrate or surface can also make a strong contribution to 1/*f* noise. An ability to change the thickness of few-layer graphene (FLG) conductors by one atomic layer at a time opens up opportunities for examining surface and volume contributions to 1/*f* noise directly.

## I.      Importance of the 1/*f* Noise for Graphene Applications

In addition to the scientific significance of investigating 1/*f* noise in a 2D system, there are practical reasons why 1/*f* noise characteristics of graphene are particularly important. They are related to graphene's physical properties and envisioned applications [13]. The most promising electronic applications of graphene are likely those that are not strongly hampered by the absence of the energy band gap but rather rely on graphene's exceptionally high electron mobility, $\mu$, thermal conductivity, saturation velocity, $v_S$, and the possibility of tuning the carrier concentration, $n_C$, with the gate over an exceptionally wide range. The applications that fall into this category are those in chemical and biological sensors, transparent electrodes, ultra-fast transistors for communications, optoelectronic devices, interconnect wiring, and various electrodes. Indeed, the exceptional sensitivity of graphene gas sensors has been demonstrated using the relative resistance of the graphene channels, $\Delta R/R$ [14]. It was attributed to the precise control of $n_C$ with the electrostatic gating and high $\mu$ of graphene. The prospects of high-frequency graphene transistors for communication, which rely on its high $\mu$ and $v_S$ also look promising [15-17]. The symmetry of the electron band structure and wide-range variation of the carrier density in graphene were used to increase the functionality of amplifiers and phase detectors utilized in communications and signal processing [17]. For all mentioned applications, 1/*f* noise is a crucial performance metric.

The sensitivity of amplifiers and transducers used in sensors is ultimately defined by the flicker noise level [18-19]. The accuracy of a system limited by $1/f^{\gamma}$ noise cannot be improved by extending the measuring time, $T \propto 1/f$, if $\gamma \gtrsim 1$. The energy, $E$, of a measured signal can be written as an integral of the square of its amplitude spectrum $E \propto \int (1/f^{\gamma})^2 df$ [18]. It is seen from this integral that for $\gamma \gtrsim 1$, the total accumulated energy of the flicker noise increases at least as fast as the measuring time $T$. In contrast, when measuring white noise, e.g. shot or thermal noise, the accuracy increases as $T^{1/2}$. The *sensitivity* and *selectivity* of many types of sensors, particularly those, that rely on electrical response is limited by $1/f$ noise [18-20]. The same considerations apply for graphene sensors.

Although $1/f$ noise dominates the spectrum only at low frequencies, its level is important for communications at high frequencies, because $1/f$ noise is the major contributor to the phase noise of the oscillating systems (see Box I). The phase noise of an oscillator, i.e. spectral selectivity, determines a system's ability to separate adjacent signals. The up-conversion of $1/f$ noise is a result of unavoidable non-linearities in the electronic systems, which leads to $(1/f)^3$ phase noise contributions [19]. The level of $1/f$ noise is important for determining the competitiveness of graphene technology for cell phones, radars or other communication applications. These considerations explain the practical needs for a detailed investigation of $1/f$ noise in graphene devices.

## II.      Characteristics of 1/$f$ Noise in Graphene

The first report of $1/f$ noise in graphene appeared in 2008 [21-22].  It was quickly followed by a large number of studies of $1/f$ noise in graphene and FLG devices of different configurations and under various biasing conditions [23-38]. Despite major progress in the investigation of $1/f$ noise in graphene, many issues remain the subject of considerable debate. The latter is expected from the timeline of knowledge accumulation and the understanding of $1/f$ noise in other, more conventional, materials [6]. In this section we summarize the $1/f$ noise characteristics of graphene, which can be considered commonly accepted or reproducibly measured in different laboratories.

Published reports agree that the low-frequency noise in graphene is scale invariant and reveals a $1/f$ spectral dependence with the corner frequencies, $f_c$, in the range from ~1 to 100 kHz, which is similar to metals and semiconductors [21-36]. Figure 1 (a-f) shows typical $1/f$ noise characteristics of graphene devices. In a few instances generation-recombination (G-R) type bulges were observed in the low-frequency noise spectrum [23]. They were attributed to defects on the edges of graphene channels, with some characteristic times constants, which dominated the fluctuations. The noise spectral density $S_I$ is proportional to $I^2$ in graphene. The latter implies that the current does not drive the fluctuations, but merely makes the fluctuations in the sample visible via Ohm's law [7]. Measurements of $1/f$ noise in graphene devices with large variation of the channel area, $W \times L$ ($W$ is the width and $L$ in the length), from ~ 1 to 80 $\mu m^2$, confirmed that $1/f$ noise mostly originates from graphene itself and is not dominated by metal contact contributions [36].

Together with the normalized noise spectral density, $S_V/I^2$, one can use the noise amplitude, $A = (1/N)\Sigma_{m=1}^{N} f_m S_{I_m} / I_m^2$, to characterize $1/f$ noise level (here $S_{I_m}$ and $I_m$ are the noise spectral density and drain-source current measured at $m$ different frequencies $f_m$). This definition helps to reduce measurement error at specific frequencies [21-22]. The measurements of $1/f$ noise in graphene revealed that its amplitude is relatively low [21-32]. This may appear surprising considering that graphene has the thickness of just one atomic layer and carriers in graphene are ultimately exposed to disorder and traps in the gate oxide or graphene open environment interface. Different groups reported consistent values of $S_V/I^2$ in the range from $10^{-9}$ to $10^{-7}$ Hz$^{-1}$ at $f$=10 Hz or $A$~$10^{-9} - 10^{-7}$ for $\mu m$-scale channels [21-32]. The channel area, $L \times W$, normalized noise $(S_V/I^2)(L \times W)$ is ~$10^{-8} - 10^{-7}$ $\mu m^2$/Hz for $\mu m$-scale graphene devices.

Most reports are in agreement that $1/f$ noise in graphene reveals an unusual gate bias dependence [28, 30, 32, 36-38]. Close to the Dirac point, the noise amplitude follows a V-shape dependence attaining its minimum at the Dirac point where the resistance is at its maximum (see Figure 1 (c)). This dependence was reported independently by several groups using graphene devices, which varied in their design and fabrication procedures. In some graphene devices, V-shape becomes M-shape dependence over the extended bias range [28, 36-38]. There are several proposed explanations of V and M-shape gate-bias dependence [28, 30, 32, 37]. The authors of Ref. [28] attributed M-shape dependence of the noise amplitude to the spatial charge inhomogeneity related to the presence of the electron and hole puddles in graphene. Another explanation originated from the observation that M-type behavior before annealing transformed to V-type behavior after annealing, irrespective of the changes in the mobility of the graphene samples [37]. The transformation was attributed to the interplay between the long- and short-range scattering mechanisms. Water contamination of the graphene surface was found to significantly enhance the noise magnitude and change the type of the noise behavior. Removal of water by annealing results in the suppression of the long-range scattering [37].

The unusual gate dependence of the noise amplitude in graphene observed in many experiments supports the conclusion that $1/f$ noise in graphene devices does not follow the McWhorter model conventionally used for Si CMOS devices and other metal-oxide-semiconductor field-effect transistors (MOSFETs). The McWhorter model predicts that $S_I/I^2$ decreases in the inversion regime as $\sim(1/n_C)^2$, where $n_C$ is the channel carrier concentration [36, 39-40]. Any deviation from this behavior is interpreted as the influence of the contacts, inhomogeneous trap distribution in energy or space or contributions of the mobility fluctuations to the noise [39-40]. Figure 1 (e) shows the McWhorter model predictions for the normalized noise amplitudes calculated for different trap concentrations. The regions between lines 1 and 2 and between lines 2 and 3 correspond to the typical noise levels in regular Si n-channel MOSFETs and in Si MOSFETs with high-k dielectric, respectively [36]. The shaded region between horizontal lines represents the results for the noise spectral density measured in graphene FETs. With a large $n_C$, noise in graphene is higher than in typical Si MOSFETs, while a small $n_C$ yields a noise level in graphene FETs that is lower than in Si MOSFETs. The latter is despite the immature state of graphene technology compared to Si CMOS technology.

A recent study explained the observed carrier density dependence of $1/f$ noise in graphene within the mobility fluctuation approach (using an expression originating from Eq. (B2)) and taking into account the gate-bias dependence of the electron mean free path, $\Lambda$, and the scattering cross sections $\sigma_1$ and $\sigma_2$ of the long-range and short-range scattering centers [41]. An independent investigation of $1/f$ noise in a wide selection of graphene devices ($\mu$ in the range from 400 to 20000 cm$^2$/Vs) concluded that in most of their examined devices the dominant contribution to $1/f$ noise was from the mobility fluctuations arising from the fluctuations in the scattering cross section $\sigma$ [38]. The authors termed this noise mechanism "configuration noise" with the noise density proportional to $\Lambda^2 \sigma^2$ [38]. The latter suggests a similarity between these approaches and consistency with Eq. (B2). One should note that the carrier number and mobility fluctuation mechanisms can be closely related since the fluctuation in the scattering cross sections $\sigma$ of the scattering centers can be due to the capture or emission of electrons, which also changes $N$.

The $1/f$ noise dependence on the number of atomic planes, $n_A$, in FLG devices can shed light on the physical mechanism of $1/f$ noise. It is also important for practical applications. Increasing $n_A$ reduces the electron mobility and complicates gating. The benefits of a larger $n_A$ in FLG include larger currents and a weaker influence of traps inside the gate dielectrics on the electron transport inside FLG channel. It was reported that the noise in bilayer graphene (BLG) channels is lower than in single-layer graphene (SLG) [21]. The authors suggested that $1/f$ noise reduction in BLG is associated with its band structure that varies with the charge distribution between the two atomic planes resulting in screening of the potential fluctuations owing to the external impurity charges [21]. It was later confirmed that $1/f$ noise level continues to decrease with increasing thickness of FLG conductors. Figure 1 (f) shows the experimentally determined trend for noise reduction with increasing number of the atomic planes, $n_A$, i.e. the channel thickness $H = n_A \times h$, where $h = 0.35$ nm is the thickness of SLG.

The volume noise originated from independent fluctuators scaled inversely proportional to the sample volume. Therefore for the constant area film noise is inversely proportional to its thickness $H$, $S_I/I^2 \propto 1/H$. Such dependence observed experimentally can be interpreted as an indication of a volume noise mechanism [9, 42]. If noise originates from the surface, varying the thickness of the film serves only to change the fraction of the current passing through the surface layer. Then the $1/f$ noise would depend on the thickness according to $S_I/I^2 \propto (1/H)^2$ [12, 43]. Previous attempts to test directly whether $1/f$ noise is dominated by contributions coming from the sample surface or its volume have not led to conclusive answer because of inability to fabricate continuous metal or semiconductor films with the uniform thickness below ~8 nm [12]. Unlike the thickness of metal or semiconductor films, the thickness of FLG can be continuously and uniformly varied all the way down to a single atomic layer of graphene – the actual *surface*. It was recently found that $1/f$ noise in FLG becomes dominated by the volume noise when the thickness exceeds $n_A$~7 (~2.5 nm) [44]. The $1/f$ noise is the surface phenomenon below this thickness. At the high-bias regime, the surface contributions are more pronounced even for larger *H* [44].

### III. Noise Reduction in Graphene Devices

As indicated above the noise amplitudes of $\sim 10^{-9} - 10^{-7}$ reported for µm-size graphene channels are relatively low. A comparison with carbon nanotubes shows that graphene devices have lower resistance and about three orders of magnitude smaller noise amplitude [45]. Environmental exposure and aging increased the level of 1/*f* noise [36]. Deposition of the top-dielectric in the top-gate graphene FETs results in mobility degradation but does not substantially increase the noise level [24]. The latter suggest that the use of the high-quality cap layers on top of graphene channels may prevent 1/*f* noise increase under environmental exposure. Practical applications of graphene, particularly in low-power devices with nm-scale channels, will require further reduction in 1/*f* the noise level. It is generally true that as the technology matures, the level of 1/*f* noise decreases [6]. A smaller density of structural defects and higher material quality usually results in smaller noise spectral density. Special processing steps or device designs can lead to substantial reduction in the noise level. For example, it was shown that GaN/AlGaN heterostructure field-effect transistors (HFETs) where the high current density is achieved via increasing Al content in the barrier layer – the so-called "piezo-doping" – reveal lower 1/*f* noise level than GaN/AlGaN HFETs with conventional channel doping [46]. Several possible methods of 1/*f* noise reduction in graphene FETs have also been reported.

In one approach, the device channel was implemented with FLG with the thickness varied from SLG in the middle to BLG or FLG at the source and drain contacts (Figure 2 (a-b)). It was found that such graphene thickness-graded (GTG) devices have *µ* comparable to the reference SLG devices while producing lower noise levels [47]. The electron density of states (DOS) in SLG in the vicinity of its Dirac point is low owing to the Dirac-cone linear dispersion. Even a small amount of the charge transfer from or to the metal can strongly affect the Fermi energy of graphene. The values of $\Delta E_F$=-0.23 eV and $\Delta E_F$=0.25 eV were reported for Ti and Au contacts to graphene, respectively [48]. The quadratic energy dispersion in BLG or FLG results in DOS, which is different from that in graphene. The same amount of charge transfer determined by the work function difference will lead to the smaller Fermi level shifts in BLG and FLG than in single-layer graphene owing to the larger DOS in BLG and FLG (see inset to Figure 2 (a)). The potential barrier fluctuations will be smaller at the metal-BLG or metal-FLG interface than in the metal-SLG interface, resulting in lower noise level [47].

Another approach is related to the electron irradiation treatment of graphene channels [49]. It was recently reported that $1/f$ noise in graphene reveals an interesting characteristic − it reduces after irradiation (see Figure 2 (c-d)). It was experimentally observed that bombardment of graphene devices with the low-energy 20-keV electrons, which induce defects but do not eject carbon atoms, can reduce $S_V/I^2$ by an order-of magnitude at a radiation dose of $10^4$ μC/cm$^2$ [49]. It was indicated that noise reduction in graphene under irradiation can be more readily interpreted within the mobility fluctuation model. The electron beam irradiation may not produce a major change in the number of scattering centers $N_t^\mu$ contributing to $1/f$ noise while strongly reducing the electron mobility, $\mu$ and, correspondingly, mean free path $\Lambda$ leading to the reduced $1/f$ noise level (see Eq. (B2) in the Box I). In graphene, mobility is limited by the long-range Coulomb scattering from charged defects even at RT, in contrast to semiconductors or metals, where the RT mobility is typically limited by phonons, even if the defect concentration is high. The latter can explain why the effect produced by electron irradiation on $1/f$ noise in graphene differs from that in conventional materials. The noise reduction comes at the expense of mobility degradation. However, this trade-off is feasible since $\mu$ after irradiation still remains sufficiently high for practical applications.

## IV.    Challenges and Opportunities

The field of $1/f$ noise in graphene is still far from being mature. It experiences a surge in the number of experimental reports and various models proposed for explanation of particular aspects of $1/f$ noise in graphene. The challenges that have to be addressed to facilitate development of graphene technology are the following. First, there is a need in the theory, which would explain the unusual gate bias dependence of $1/f$ noise in graphene. The developed theoretical models can be incorporated in computer-aided design tools used for graphene device structure optimization. Second, the influence of metal contacts, surface contamination or analyte molecules attached to graphene channels on the low-frequency noise characteristics have to be closely examined. Considering that the electronic applications and fabrication of sensor arrays require nm-scale devices the third important challenge would be to understand what happens with $1/f$ noise when graphene channels' length and width are on the nm-length scale. It was established for conventional Si CMOS technology that the average $1/f$ noise level exhibits a much stronger than linear increase upon reducing the device size [50]. The initial report of $1/f$ noise in graphene nanoribbons [51-52] found increased noise amplitude, $A \sim 10^{-6} \div 10^{-5}$, for nanoribbons with the width of $\sim 40 \div 70$ nm [51]. It was also suggested that the conductance fluctuations are correlated with the electron DOS revealing peaks in the noise spectral density with the positions matching the electron subband energies [51-52]. In the devices where the width of graphene channels scales down to just a few nanometers one may need to consider the electron hopping transport regime and corresponding implications for $1/f$ noise. It is known that the level of $1/f$ noise in the "hopping" conductors increases with decreasing temperature [53-54], which is opposite to what is normally observed in regular conductors. Finally, variability effects in graphene, originating from environmental disturbance and material and process variations [55] have to be studied systematically and separated from the fundamental noise characteristics.

Although detrimental in many of its manifestations, low-frequency noise presents opportunities for materials characterization and can serve positive functions when used cleverly. The low-frequency noise spectroscopy can provide information about the trap levels and charge carrier dynamics. It was also used to detect degradation in interconnects. The low-frequency noise in graphene is no exception (see Figure 3). It was reported that the use of the noise spectral density, $S_I/I^2$, together with the resistance change $\Delta R/R$ in graphene sensors allows one to perform selective detection of gas molecules with graphene devices without prior functionalization of their surfaces [56]. The same approach can be extended to the label-free graphene biosensors. It is reasonable to expect more of such device concepts where excellent electronic properties of graphene are complemented by its unusual noise characteristics. In terms of fundamental science, graphene-FLG constitutes a unique material system, which allows one to investigate $1/f$ noise evolution as the dimensionality changes from bulk to 2D surface [44]. The implications of this investigation can go beyond graphene related materials. Addressing these challenges and opportunities will allow one to fully exploit graphene's potential for ultra-sensitive and selective sensors and high-speed communication applications.

**Box I: Summary of the Intrinsic Noise Types and 1/f Noise Fundamentals**

Various types of noise are commonly classified into four intrinsic noise types: (i) thermal or Johnson noise, (ii) shot noise, (iiii) generation-recombination (G-R) noise, and (iv) flicker or 1/f noise [6]. The spectral density of thermal noise is given by the Nyquist's formula $S_I(f)=4k_BT/R$, where $k_B$ is the Boltzmann's constant, $T$ is the temperature and $R$ is the resistance. The spectral density of shot noise is given by the Schottky's theorem $S_I(f)=2q<I>$, where $q$ is the charge of an electron and $<I>$ is the average value of the electrical current. Thermal and short noise types are manifestations of the random motion of charge carriers. Both noise types are called white noise because their spectral density does not depend on the frequency $f$. G-R noise is observed at low $f$ and its spectral density is described by the Lorentzian: $S_I(f)=S_0/[1+(2\pi f\tau)^2]$, where $S_0$ is the frequency independent portion of $S_I(f)$ observed at $f<f_0=(2\pi\tau)^{-1}$ and $\tau$ is the time constant associate with a specific trapping state (e.g. ionized impurity). Unlike other intrinsic noise types, 1/f noise can originate from different fluctuation processes either in the charge carrier number, mobility or both.

The most common description of 1/f noise, dominated by the carrier number fluctuations, stems from the observation that a superposition of individual G-R noise sources with the lifetime distributed on a logarithmically wide time scale, within the $\tau_1$ and $\tau_2$ limits, gives the 1/f spectrum in the intermediate range of frequencies $1/\tau_2 < \omega < 1/\tau_1$ [57]. Introducing a density distribution of lifetimes, $g(\tau_N)$, one can write the spectral density of the number fluctuations, $S_N$, in the form

$$S_N(\omega) = 4\overline{\delta N^2}\int_{\tau_1}^{\tau_2} g(\tau_N)\frac{\tau_N}{1+(\omega\tau_N)^2}d\tau_N \ . \tag{B1}$$

Integration of Eq. (B1) for $g(\tau_N)=[\tau_N\ln(\tau_2/\tau_1)]^{-1}$, gives the 1/f spectrum inside the region determined by the limiting values of $\tau_N$. Further development of this idea in the context of semiconductors led to a model – commonly referred to as McWhorter's model [58] – which is used to describe 1/f noise in conventional field-effect transistors (FETs). Consider a typical Si CMOS device structure shown in (a). Defects that act as the carrier traps are distributed inside

SiO$_2$ gate oxide layer. Each defect is characterized by its own time constant $\tau_N$, which is determined by its distance from the channel, e.g. $\tau = \tau_0 \exp(\lambda z)$, where $z$ is the distance of the trap from the channel, $\tau_0 \sim 10^{-10}$ s and $\lambda \sim 2 \times 10^8$ cm$^{-1}$ is the tunneling parameter [58-59]. Carrier capture and emission back to the channel leads to current fluctuations $\delta I \propto q(\delta N)\mu$. The contribution of traps with different $\tau$ results in a set of G-R bulges represented by Lorentzian functions. The envelope of the closely positioned Lorentzians has the $1/f$ type dependence over the relevant frequency range (b). If one type of traps dominates the fluctuation processes, e.g. traps at the interface with the same time constant, the G-R bulge associated with this trapping state can appear superimposed on the $1/f$ spectrum (c). In graphene context, G-R noise was discussed in Refs. [23, 60]. The $1/f$ spectrum reaches the white noise floor at some corner frequency $f_c$ (c). Depending on a particular device or temperature, the white noise level is defined by either thermal noise or shot noise. Specifics of shot noise in graphene were reported in Refs. [61-65]. An approach to re-cast McWhorter model of $1/f$ noise specifically for graphene was reported in Ref. [66]. It was suggested that the observed noise in graphene correlates better with the charge scattering primarily due to the long-range Coulomb scattering from charged impurities rather than short-range scattering from lattice defects [66].

The low-frequency $1/f$ noise caused by mobility fluctuations can appear as a result of the superposition of elementary events in which the scattering cross-section, $\sigma$, of the scattering centers fluctuates changing from $\sigma_I$ to $\sigma_2$. The cross-section can change owing to capture or release of the charge carriers. In the framework of the mobility-fluctuation model, the noise spectral density of the elemental fluctuation events contributing to $1/f$ noise in any material is given by [67-69]

$$\frac{S_I}{I^2} \propto \frac{N_t^{\mu}}{V} \frac{\tau \varsigma (1-\varsigma)}{1+(\omega \tau)^2} \Lambda^2 (\sigma_2 - \sigma_1)^2, \qquad (B2)$$

where $N_t^{\mu}$ is the concentration of the scattering centers of a given type that contribute to the noise, $\Lambda$ is the mean free path of the charge carriers, $\varsigma$ is the probability for a scattering center to be in the state with the cross-section $\sigma_I$. Integration of Eq. (B2) results in the $1/f$ spectrum caused by the mobility fluctuations.

The absence of a single noise mechanism complicates an introduction of a meaningful figure of merit for 1/*f* noise. The most commonly used figure of merit – Hooge parameter $\alpha_H$ – is based on his empirical formula [9]

$$S_R / R^2 = \alpha_H / Nf \, , \tag{B3}$$

where $S_R \sim (\delta R)^2$ is the power spectral density of the fluctuations in the value of the resistance ($S_R/R^2 = S_I/I^2 = S_V/V^2$). Eq. (B3) was introduced specifically for the *mobility* fluctuations but then extended to other 1/*f* noise mechanisms for the purpose of noise level comparison. The application of this figure of merit introduced for volume noise to a 2D system such as graphene presents conceptual difficulties.

Although 1/*f* noise dominates the spectrum only at low-frequency, it up-converts to high frequencies, owing to unavoidable non-linearities in the devices or systems (d). As a result, 1/*f* noise makes up the main contribution to the phase noise of communication systems and sensors (d). Downscaling of any material system for the use in nm-scale devices can further increase 1/*f* noise level and complicate practical applications [50, 70].

### *Acknowledgements*

This work was supported, in part, by the Semiconductor Research Corporation (SRC) and Defence Advanced Research Project Agency (DARPA) through FCRP Center for Function Accelerated nanoMaterial Engineering (FAME) and by the National Science Foundation (NSF) projects CCF-1217382, EECS-1128304, EECS-1124733, and EECS-1102074. The author is indebted to Prof. S. Rumyantsev (RPI and Ioffe Institute) for critical reading of the manuscript and providing valuable suggestions. He also acknowledges insightful discussions on 1/$f$ noise in graphene with Prof. M. Shur (RPI).

**FUGURES CAPTIONS**

**Figure 1: Noise characteristics of graphene devices.** (a) Normalized noise spectral density, $S_I/I^2$, of a top-gated graphene device as a function of frequency, $f$, for a range of gate biases $V_G$=0 (black), 10 V (red), 20 V (green), 30 V (blue) and 40 V (light blue). The source-drain voltage is $V_{DS}$=50 mV. The inset shows scanning electron microscopy (SEM) image of the top-gate graphene FET. (b) Noise spectral density in different graphene devices normalized by the graphene channel area $W \times L$ as a function of the gate bias, $V_G$. The data points in blues color (circles, triangles and rectangles) are for three SLG devices while the rest of the data points are for BLG devices. (c) Noise amplitude as the function of the gate bias and channel resistance in a graphene device. The data shows the V-type noise behavior consistent with many independent reports. (d) Experimental M-shape dependence of $1/f$ noise spectral density on the gate bias reported in several studies. The vertical lines indicate the carrier density $n_C \sim 10^{12}$ cm$^{-2}$. (e) Noise spectral density multiplied by the graphene channel area as a function of the gate voltage. The tilted straight lines are calculated from the McWhorter model for three different gate-oxide trap concentrations: (1) is for $N_T$=5×10$^{16}$ (cm$^3$eV)$^{-1}$, (2) is for $N_T$=10$^{18}$ (cm$^3$eV)$^{-1}$ and (3) is for $N_T$=10$^{20}$ (cm$^3$eV)$^{-1}$. The shadowed region represents the experimental noise level for graphene transistors. The frequency of the analysis is $f$=10 Hz. The data indicates that 1/f noise in graphene does not follow $(1/n_C)^2$ dependence characteristic for conventional FETs. (f) Noise spectral density, $S_I/I^2$, in FLG as a function of frequency shown for three devices with distinctly different thickness defined by the number of atomic planes $n$=1 (blue), $n \approx 7$ (red) and $n \approx 12$ (green). Figures (a), (c) and (f) are reprinted with permission from the American Institute of Physics (IOP). Figure (d) is reprinted with permission from the American Chemical Society (ACS).

**Figure 2: Noise reduction in graphene devices.** (a) Normalized noise spectral density in a typical back-gated graphene device. The inset illustrates the design of the graded-thickness graphene FET with the channel thickness gradually changing from graphene to FLG near the metal contacts. (b) Normalized noise spectral density of GTG FETs and the reference SLG and BLG FETs as the function of the graphene channel area. The filled symbols represent SLG, the open symbols – BLG while the half-filled symbols indicate the data-points for GTG FETs. For each device the noise level is shown for several biasing points within the $|V_G$-$V_D|\leq30$ V range from the Dirac point $V_D$. Noise increases as bias points shift away from $V_D$. The dashed lines are given as guides to the eye. Note that GTG FETs have a comparably reduced noise level to that in BLG FETs, while revealing an electron mobility that is almost as high as in graphene FETs. The inset shows the band structures of SLG with the linear dispersion and BLG with the parabolic dispersion the vicinity of the charge neutrality point. (c) Normalized noise spectral density as a function of frequency for a graphene device after each irradiation step. The source-drain bias was varied from 10 mV to 30 mV. The date before irradiation marked as BR. Note that $1/f$ noise decreases monotonically with increasing irradiation dose indicated as RD. (d) Normalized noise spectral density as a function of the radiation dose at zero gate bias. The arrows indicate the level of $1/f$ noise before irradiation. The Figures (a), (b), (c) and (d) are reprinted with permission from the American Institute of Physics (IOP).

**Figure 3: Low-frequency noise as a sensing signal.** (a) Normalized noise spectral density $S_I/I^2$ multiplied by frequency $f$ versus frequency $f$ for the device in open air and under the influence of different vapors. Different vapors induce noise with different characteristic frequencies $f_c$. The frequencies, $f_c$, are shown explicitly for two different gases. The solid lines show the polynomial fitting of the experimental data. (b) Normalized noise spectral density multiplied by frequency $f$ versus frequency $f$ for three different graphene devices exposed to acetonitrile vapor. Note the excellent reproducibility of the noise response of the graphene devices showing the same frequency $f_c$ for all three devices. The inset presents SEM image of the label-free graphene sensor. The scale bar is 3 μm. The Figures (a) and (b) are reprinted with permission from the American Chemical Society (ACS).

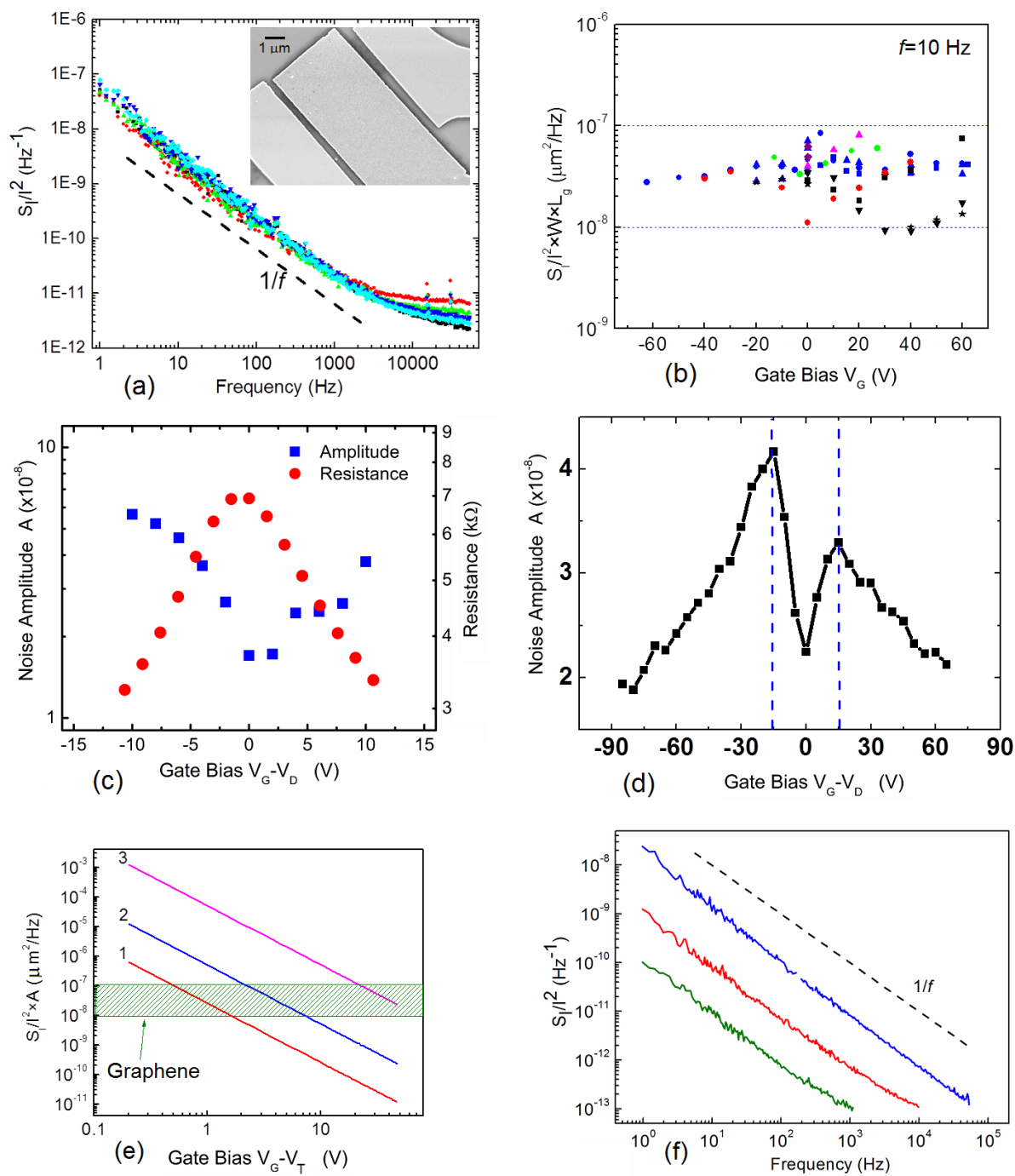

Figure 1

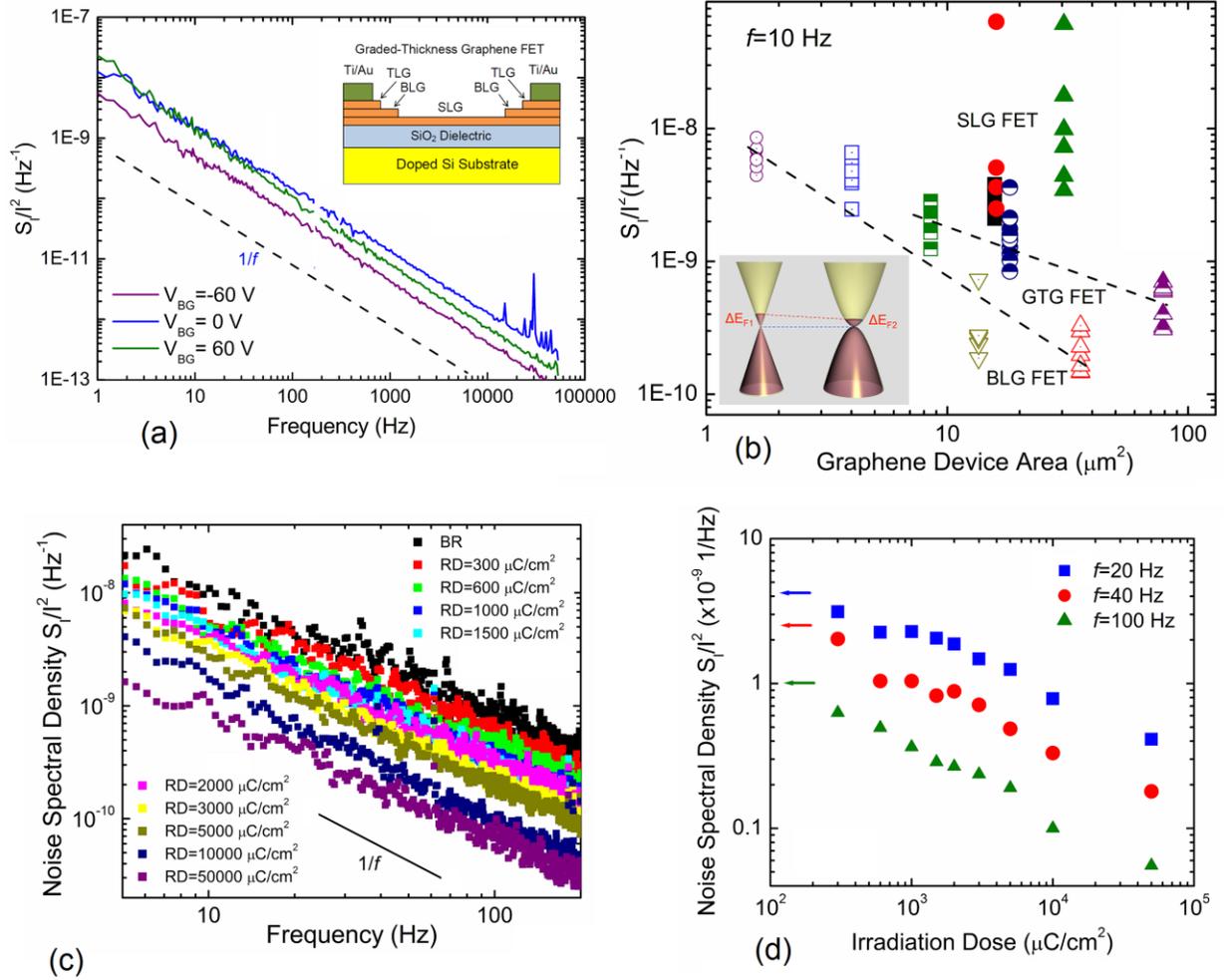

Figure 2

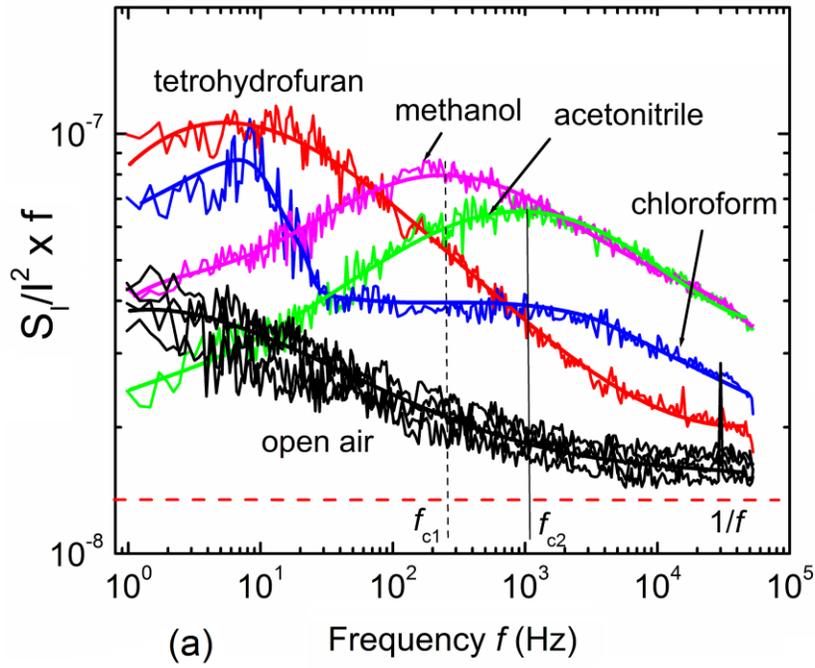

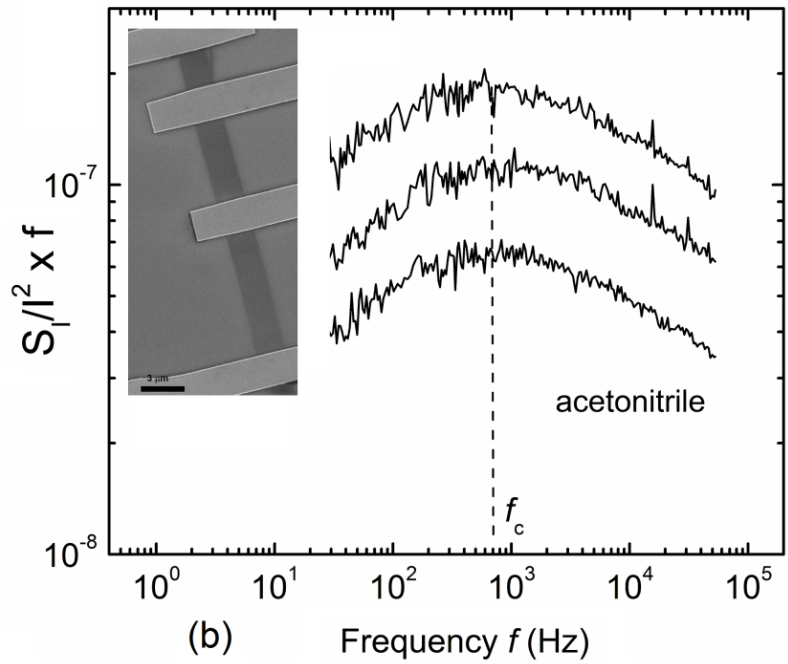

Figure 3

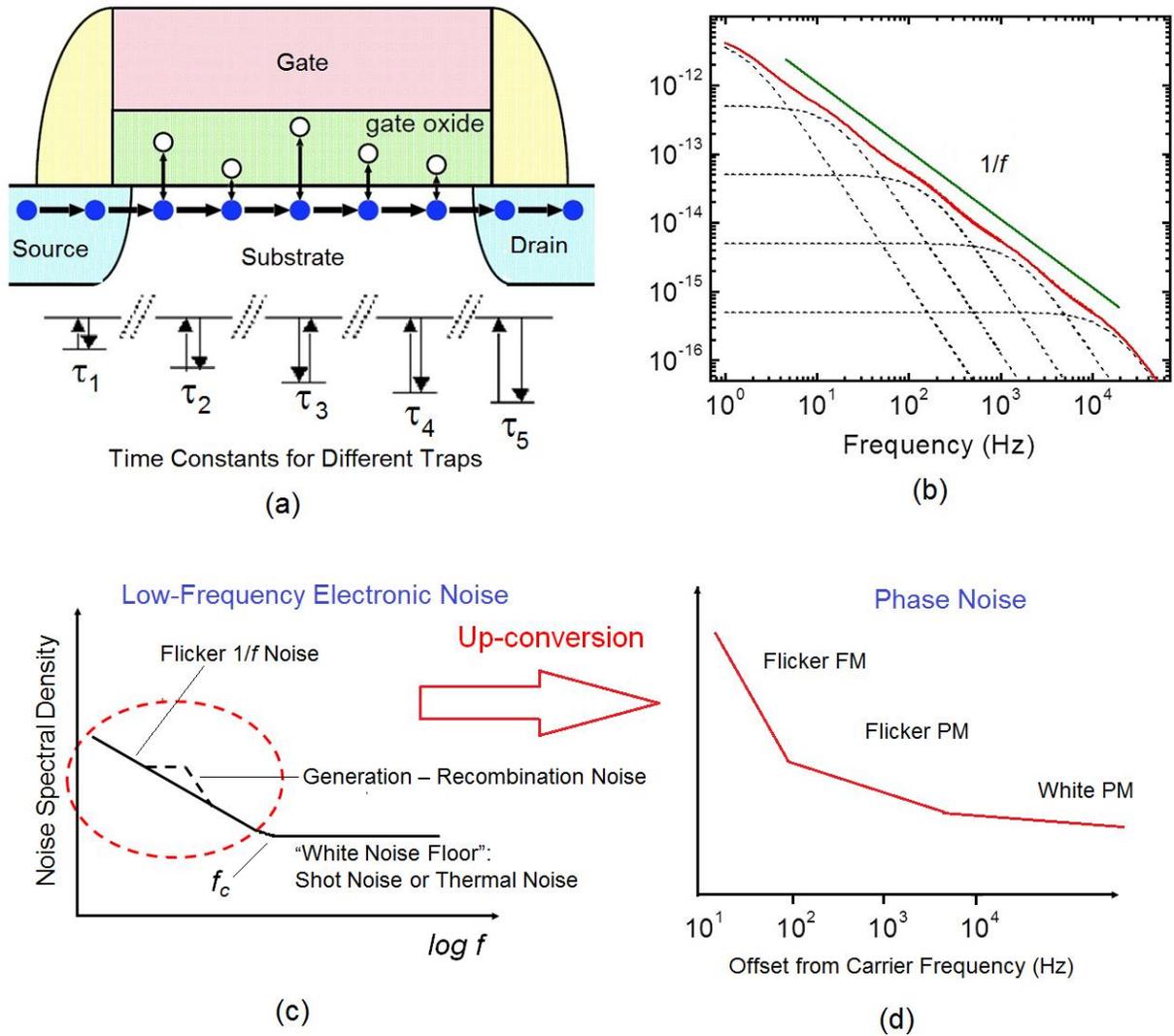

(a)

(b)

(c)

(d)

Figure for the Text Box